\documentclass[aps,prl,twocolumn,superscriptaddress,floatfix]{revtex4-1}
\usepackage{amssymb}

\usepackage{amsmath}
\usepackage{graphicx}
\usepackage{dcolumn}
\usepackage{bm}

\begin{document}

\title{Dynamics of Field Induced Polarization Reversal in Strained
Perovskite Ferroelectric Films with c-oriented Polarization}
\author{Laurent Baudry}
\email{laurent.baudry@iemn.univ-lille1.fr}
\affiliation{ Institute of Electronics, Microelectronics and Nanotechnology (IEMN)-DHS D\'epartment, UMR CNRS 8520, Universit\'e des Sciences et Technologies de Lille, 59652 Villeneuve d'Ascq Cedex, France}
\author{Igor A. Luk'yanchuk}
\affiliation{Laboratory of Condensed Matter Physics, University of
Picardie Jules Verne, Amiens, 80039, France} \affiliation{L. D.
Landau Institute for Theoretical Physics, Moscow, Russia}
\author{Anna Razumnaya}
 \affiliation{Physics
Department, Southern Federal University, Rostov on Don, 344090
Russia }

\date{\today }

\begin{abstract}
The field-induced polarization reversal in $c$-oriented
ferroelectric phase of strained perovskite film has been studied. We
show that in additional to the conventional longitudinal switching
mechanism, when c-oriented polarization vector changes its modulus,
the longitudinal-transversal and transversal mechanisms when the
perpendicular component of polarization is dynamically admixed are
possible. The later process can occurs either via the
straight-abrupt or initially-continues polarization turnover
scenario. We specified the obtained results for the case of
PbTiO$_{3}$ and BaTiO$_{3}$ ferroelectrics and propose the
experimental methods for their investigation.
\end{abstract}

\pacs{77}
\maketitle

\begin{figure}[!h]
\centering
\includegraphics[width=0.34\textwidth]{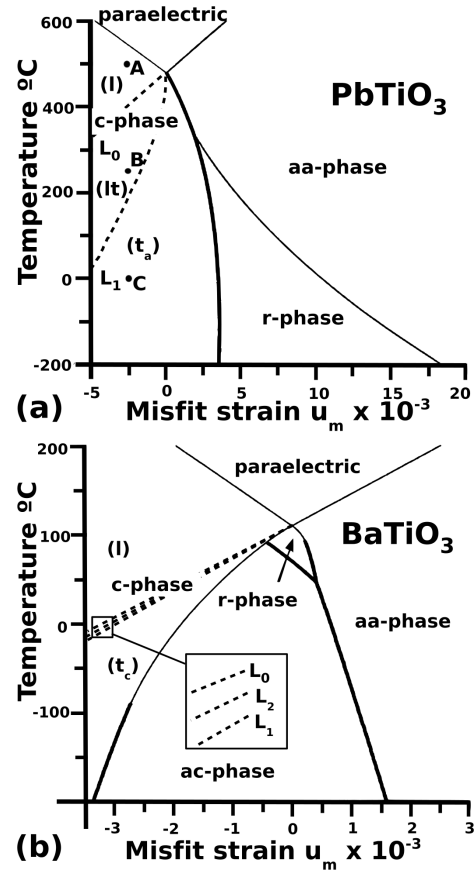}
\caption{
Regions of the longitudinal (\emph{l}),
 longitudinal-transversal (\emph{lt})
and transversal (\emph{t}) switching regimes and
corresponding separating lines L$_{0}$ and L$_{1}$  on the phase
diagrams of strained films of PbTiO$_{3}$ (a) and BaTiO$_{3}$ (b),
adopted from Ref. \protect\cite{PERTSEV_1998}. Stability line
L$_{2}$ determines the type of transversal switching. Close location
of L$_{2}$ and L$_1$ for BaTiO$_{3}$  implies that it occurs
according the initially-continuous turnover of polarization ($t_c$),
whereas the absence of this line for PbTiO$_{3}$ means that
transversal switching is straight-abrupt ($t_a$)
} \label{DIAG}
\end{figure}

Dynamical switching properties of ferroelectrics are essential for their
application in the memory-storage devices \cite{SCOTT_2000}. The underlying
mechanism of polarization reversal is of special interest for the mostly
used pseudo-cubic perovskite crystals that, depending on the orientation of
polarization $\mathbf{P}=(P_{1},P_{2},P_{3})$ can exhibit tetragonal,
orthorhombic or rhombohedral structural phases in the ferroelectric state of
the bulk material \cite{LINES_1977}. The situation is more diverse in case
of substrate-deposited perovskite ferroelectric films in which the
substrate-provided deformation makes the lattice constant $c$ in $z$%
-direction (perpendicular to the film surface) different from the
in-plane lattice constants $a=b$ already in the high-temperature
paraelectric phase with $\mathbf{P}=0$. In particularly, Pertsev,
Zembilgotov and Tagantsev \cite{PERTSEV_1998,PERTSEV_1999} studied
the effect of substrate clamping on PbTiO$_{3}$ and BaTiO$_{3}$
films and proposed that at least four structural
phases can exist in strain-temperature, $u_{m}$-$T$ phase diagram (Fig.~\ref%
{DIAG}). The so-called $c$-phase with $\mathbf{P}=(0,0,P_{3})$ occurs at
high compressive strains whereas the $aa$-phase with $\mathbf{P}%
=(P_{1},P_{1},0)$ is realized at high tensile strains. Either $ac$-phase
with $\mathbf{P}=(P_{1},0,P_{3})$ or $r$-phase with $\mathbf{P}%
=(P_{1},P_{1},P_{3})$ can occur at low strains. These phases are
thermodynamically stable and separated by continuous (thin) or discontinuous
(bold) transition lines in Fig.~\ref{DIAG}.

In the present letter we study the uniform polarization switching in PbTiO$_{3}$
and BaTiO$_{3}$ oxides induced by the applied electric field and
demonstrate that the situation is even more rich. Additional phases
can dynamically
appear during the polarization reversal. We restrict ourselves to the $c$%
-phase region of $u_{m}$-$T$ phase diagram and consider the switching
process when\ the initially up-oriented polarization $\mathbf{P}=(0,0,P_{3})$
decreases and then suddenly drops down under the oppositely applied field $%
\mathbf{E}=(0,0,E)$ with $E<0$.

To describe the PbTiO$_{3}$ and BaTiO$_{3}$ materials we use the \
renormalized Landau-Devonshire functional given in
\cite{PERTSEV_1998}, for which the account of the six-order terms is
known to be important \cite{BELL_1984,LI_2005,WANG_2007}:
\begin{eqnarray}
&&\tilde{G}\left( \mathbf{P},E,T,u_{m}\right) =a_{1}^{\ast }\left(
P_{1}^{2}+P_{2}^{2}\right) +a_{3}^{\ast }P_{3}^{2}+a_{11}^{\ast }\left(
P_{1}^{4}+P_{2}^{4}\right)  \notag \\
&&+a_{33}^{\ast }P_{3}^{4}+a_{13}^{\ast }\left( P_{1}^{2}+P_{2}^{2}\right)
P_{3}^{2}+a_{12}^{\ast }P_{1}^{2}P_{2}^{2}+a_{123}P_{1}^{2}P_{2}^{2}P_{3}^{2}
\notag \\
&&+a_{112}\left[ P_{1}^{4}\left( P_{2}^{2}+P_{3}^{2}\right) +P_{3}^{4}\left(
P_{1}^{2}+P_{2}^{2}\right) +P_{2}^{4}\left( P_{1}^{2}+P_{3}^{2}\right) %
\right]  \notag \\
&&+a_{111}\left( P_{1}^{6}+P_{2}^{6}+P_{3}^{6}\right) +\frac{u_{m}^{2}}{%
s_{11}+s_{12}}-EP_{3}.  \label{Fun}
\end{eqnarray}%
The last term in Eq.~(\ref{Fun}) presents the field-driving interaction with
electric polarization. The renormalized coefficients $a_{1}^{\ast }$, $%
a_{3}^{\ast }$, $a_{11}^{\ast }$, $a_{33}^{\ast }$, $a_{13}^{\ast }$ and $%
a_{12}^{\ast }$ depend on the misfit strain $u_{m}$ and temperature $T$
whereas other coefficients $s_{11}$, $s_{12}$, $a_{111}$, $a_{112}$ and $%
a_{123}$ correspond to\ its bulk homologous, as was explicitly
specified in Ref.~\cite{PERTSEV_1998}.

 Note that several alternative
approaches were
proposed to establish the $u_{m}$-$T$ phase diagram of BaTiO$_{3}$ \cite%
{DIEGUEZ_2004,LAI_2005,Shirokov_2007}. Their results are competitive with
\cite{PERTSEV_1998,PERTSEV_1999} mostly in relative location of $r$- and $ac$%
- phases. This minor difference is not essential for our consideration and
can be easily taken into account for each particular case. In what follows,
we consider the competition between the switching-induced $ac$ and $\ r$
phases. By substitution of the corresponding order parameters $\mathbf{P}%
=(P_{1},0,P_{3})$ and $\mathbf{P}=(P_{1},P_{1},P_{3})$ in (\ref{Fun}) we
obtain the following effective functional:
\begin{gather}
\tilde{G}=\frac{b_{1}}{2}P_{1}^{2}+\frac{b_{3}}{2}P_{3}^{2}+\frac{b_{11}}{4}%
P_{1}^{4}+\frac{b_{33}}{4}P_{3}^{4}+\frac{b_{13}}{2}P_{1}^{2}P_{3}^{2}
\label{Fun1} \\
+\frac{b_{113}}{2}P_{1}^{4}P_{3}^{2}+\frac{b_{133}}{2}P_{1}^{2}P_{3}^{4}+%
\frac{b_{111}}{6}P_{1}^{6}+\frac{b_{333}}{6}P_{3}^{6}-EP_{3},  \notag
\end{gather}%
where $b_{1}=2a_{1}^{\ast }$, $b_{3}=2a_{3}^{\ast }$, $b_{11}=4a_{11}^{\ast
} $, $b_{13}=2a_{13}^{\ast }$, $b_{33}=4a_{33}^{\ast }$, $b_{111}=6a_{111}$,
$b_{113}=2a_{112}$, $b_{133}=2a_{112}$, $b_{333}=6a_{111}$ for $ac$-phase
case and $b_{1}=4a_{1}^{\ast }$, $b_{3}=2a_{3}^{\ast }$, \ $%
b_{11}=8a_{11}^{\ast }+2a_{12}^{\ast }$, $b_{13}=4a_{13}^{\ast }$, $%
b_{33}=4a_{33}^{\ast }$, $b_{111}=12a_{111}+12a_{112}$, $%
b_{113}=2a_{123}+4a_{112}$, $b_{133}=4a_{112}$, $b_{333}=6a_{111}$ for $r$%
-phase case.

Our approach is inspired by that given by Iwata and Ishibashi \cite%
{IWATA_1999} for the case of cubic (unstrained) lattice in paraelectric
phase. It was demonstrated that depending on the strength of the
polarization-lattice coupling, two reversal mechanisms are possible.\ For
strong cubic anisotropy the switching occurs like in uniaxial one-component
ferroelectrics by dynamical change of the modulus of the longitudinal
polarization component $P_{3}$. For weak anisotropy the transversal
component $P_{1}$ virtually admixes to $P_{3}$ during the process. Such%
\textit{\ polarization-rotation} scenario can, for instance, occurs in PbZr$%
_{x}$Ti$_{1-x}$O$_{3}$ compounds when the anisotropic coupling is soften
just as the composition parameter $x$ approaches the morphotropic point $%
x\simeq 0.44$ from above.

The distinguishing feature of the substrate-deposited films from the bulk
cubic case is the strain-induced uniaxial anisotropy that is reflected both
by the splitting of the critical temperatures in the second order $P_{1}^{2}$
and $P_{3}^{2}$ terms and by accounting for the 6th-order cross-coupling
terms. To understand the dynamical mechanism of polarization reversal we
should catch the critical field at which the switching instability occurs.
Application of an opposite electric field leads to the decrease of $c$%
-oriented polarization which stays yet positive until the critical field is
reached. \ At this stage the\ field-driven polarization evolution, $P_{3}(E)$
is given by the one-component variational equation:%
\begin{equation}
\left( \frac{\partial \tilde{G}}{\partial P_{3}}\right)
_{P_{1,2}=0}=b_{3}P_{3}+b_{33}P_{3}^{3}+b_{333}P_{3}^{5}-E=0.  \label{GL6}
\end{equation}

The value of the critical field at which polarization switching starts can
be obtained from the loss of the positive definiteness of the Hessian matrix
$H_{ij}=\frac{\partial ^{2}\tilde{G}}{\partial P_{i}\partial P_{j}}$,
presented in the extremal point of initial equilibrium $P_{1}=0$, $%
P_{3}=P_{3}(E)$ as:
\begin{gather}
H_{33}=b_{3}+3b_{33}P_{3}^{2}+5b_{333}P_{3}^{4},  \label{I1} \\
H_{11}=b_{1}+b_{13}P_{3}^{\,2}+b_{133}P_{3}^{\,4},  \label{I2} \\
H_{13}=H_{31}=0.  \label{I3}
\end{gather}%
where the dependence $P_{3}(E)$ is given by Eq.(\ref{GL6}). Upon field
increase the competition occurs between the longitudinal and transversal
critical fields $E^{(l)}$ and $E^{(t)}$, determined by the conditions $%
H_{33}\left( P_{3}\left( E^{(l)}\right) \right) =0$ and $H_{11}\left(
P_{3}\left( E^{(t)}\right) \right) =0$. Importantly, the switching occurs at
the instability field $E^{(l)}$ or $E^{(t)}$ which is attained first and the
further scenario of polarization vector evolution is determined by the
occurring type of instability.

(i) For $\left\vert E^{(l)}\right\vert <\left\vert E^{(t)}\right\vert $ the
\textit{longitudinal} (\textit{l}) switching instability is realized first
and the polarization vector reverses its direction by change of the
amplitude of $P_{3}$ from positive to negative, passing through $P_{3}=0$.

(ii) For $\left\vert E^{(t)}\right\vert <\left\vert E^{(l)}\right\vert $ the
\textit{transversal} (\textit{t}) switching instability is realized first
and the component $P_{1}$ is admixed to $P_{3}$ after the beginning of the
reversal process, just above $E^{(t)}$. Polarization switching has therefore
the rotational constituent, like in the Iwata and Ishibashi model.

(iii) There can exist also the mixed \textit{longitudinal-transversal (lt) }%
regime when the polarization reversal starts according to longitudinal
scenario at $E=E^{(l)}$ but the transversal component $P_{1}$ virtually
appears at the later stage of the process.

\begin{figure}[t]
\centering
\includegraphics[width=7.5cm]{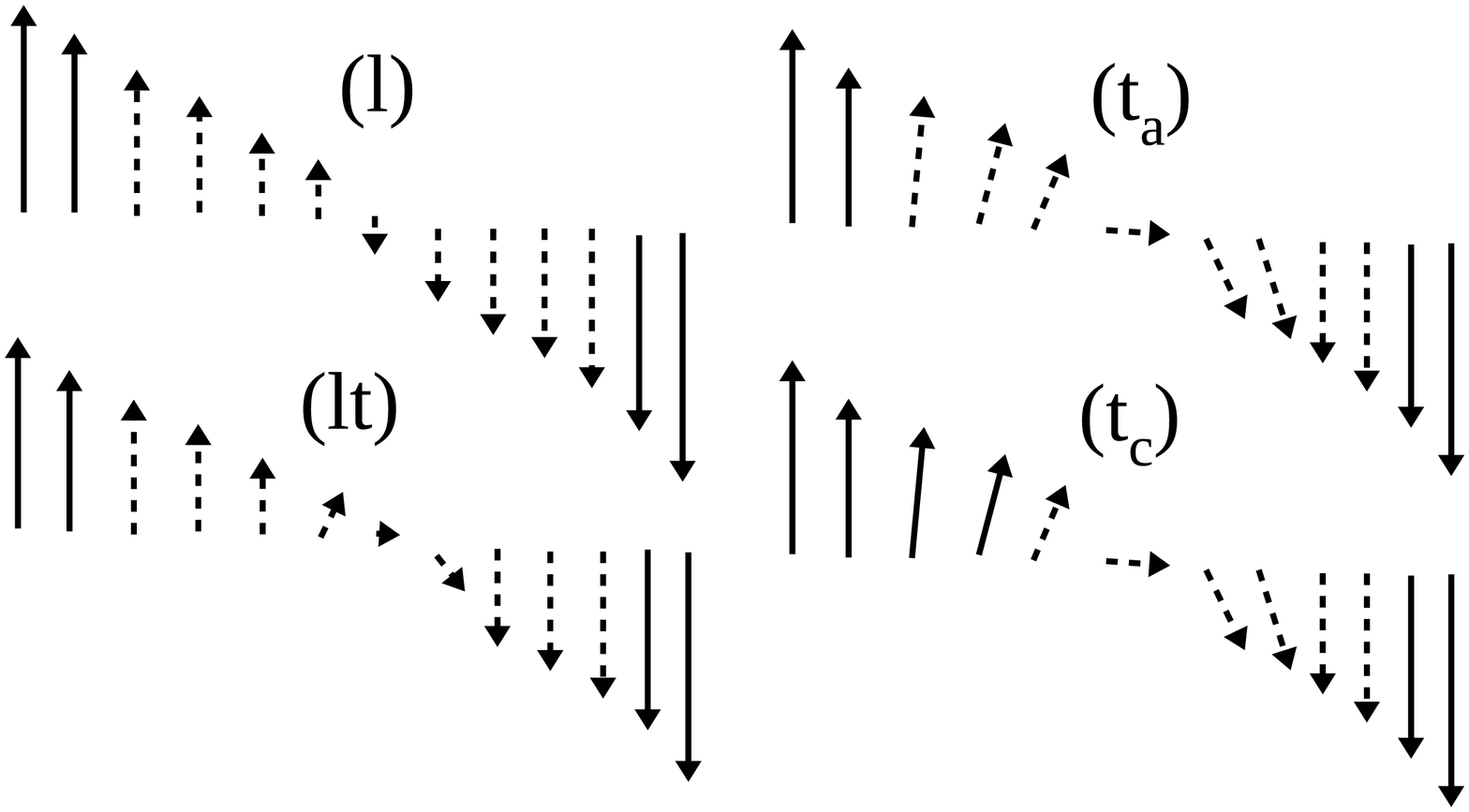}
\caption{Reversal of the polarization vector as function of
increasing with time switching electric field during longitudinal
($l$), longitudinal-transversal ($lt$),  transversal straight-abrupt
($t_a$) and transversal initialy-continuous ($t_c$) switching.
Solid lines present the thermodynamically stable field-induced
states whereas the dashed lines present the dynamically-virtual
states appearing during the abrupt switching process.} \label{DES}
\end{figure}

The polarization evolution in $l$, $lt$ and $t$ switching regimes is
sketched in Fig.~\ref{DES}. We presume that they are separated by crossover
lines L$_{0}$ and L$_{1}$ in $u_{m}$-$T$ phase diagram and find the
condition of their existence. The $t$-type switching can have either
\emph{initially-continous} ($t_{c}$) or \emph{straight-abrupt} ($t_{a}$) character as will
be specified later.

According to the given above consideration the transversal component $P_{1}$
can dynamically admix to the component $P_{3}$ during polarization reversal
if the polarization-dependent Hessian matrix element $H_{11}$ becomes
negative in course of the switching. This occur e.g., when coefficient $%
b_{1} $ is negative. Then, when the dropping-down polarization goes through the
state with vanishing $P_{3}$, the element $H_{11},$ according Eq. (\ref{I2})
acquires the negative sign in the vicinity of $P_{3}=0$. The polarization
vector will experience the instability towards the transversal deviation and
the $lt$ regime will be realized. Therefore the crossover line L$_{0}$
between $l$ and $lt$ regimes is given by the condition:
\begin{equation}
\text{L}_{0}:\quad b_{1}\left( u_{m},T\right) =0.  \label{L0}
\end{equation}%
Noteworthy that the line L$_{0}$ can be found in $u_{m}$-$T$ phase diagram
as the prolongation of the paraelectric $aa$ phase transition line located
in $u_{m}>0$ region into the $u_{m}<0$ region.
\begin{figure}[h]
\centering
\includegraphics[width=0.29\textwidth]{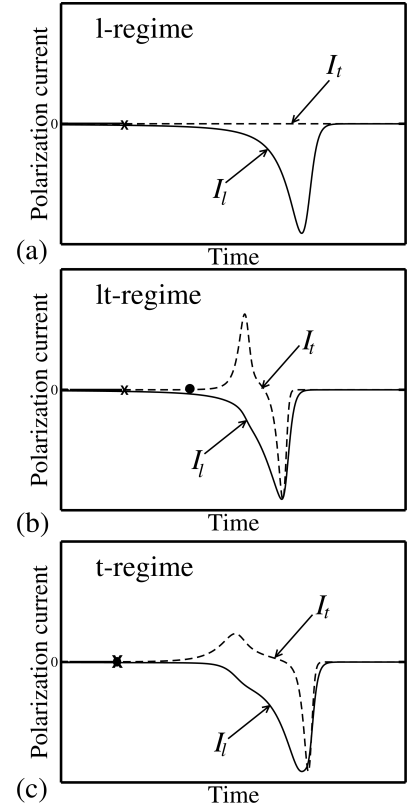}
\caption{Time dependence of the longitudinal, $I_{l}=\frac{\mathrm{d}P_{3}}{%
\mathrm{d}t}$ and transversal,
$I_{t}=\frac{\mathrm{d}P_{1}}{\mathrm{d}t}$ polarization currents
for (a) transversal (\emph{t}), (b) longitudinal-transversal
(\emph{lt}) and (c) longitudinal (\emph{l}) switching regimes for
PbTiO$_{3}$. Panels (a), (b) and (c) correspond to the points A, B
and C in Fig.~\protect\ref{DIAG}~(a). The cross and circle markers
indicate the beginning of the longitudinal and transversal
polarization reversal process correspondingly.} \label{COU}
\end{figure}

The condition of crossover between $lt$ and $t$ switching regimes can be
found by equating the critical fields $E^{(l)}$ and $E^{(t)}$ or, what is
equivalent and easier, by equating the corresponding longitudinal and
transversal critical polarizations $P_{3}^{(l)}=P_{3}\left( E^{(l)}\right) $%
, $P_{3}^{(t)}=P_{3}\left( E^{(t)}\right) $ calculated at these fields. The
latter can be found from Eqs. $H_{33}\left( P_{3}^{(l)}\right) =0$ and $%
H_{33}\left( P_{3}^{(t)}\right) =0$ as:%
\begin{eqnarray}
P_{3}^{(l)\,2} &=&\frac{\left( 9b_{33}^{2}-20b_{3}b_{333}\right)
^{1/2}-3b_{33}}{10b_{333}},  \label{Pl6} \\
P_{3}^{(t)\,2} &=&\frac{\left( b_{13}^{2}-4b_{1}b_{133}\right) ^{1/2}-b_{13}%
}{2b_{133}}.  \label{Pc6}
\end{eqnarray}%
Condition $P_{3}^{(l)}=P_{3}^{(t)}$ determines the crossover line L$_{1}$
between $lt$ and $t$ regimes:%
\begin{equation}
\text{L}_{1}\text{:}\quad \frac{b_{3}b_{13}-3b_{1}b_{33}}{%
5b_{1}b_{333}-b_{3}b_{133}}=\frac{5b_{1}b_{333}-b_{3}b_{133}}{%
3b_{33}b_{133}-5b_{13}b_{333}}.  \label{GL1}
\end{equation}

To be more specific we delimit the location of $l$, $lt$ and $t$ switching
regimes and corresponding crossover lines L$_{0}$ and L$_{1}$ on phase
diagram of strained PbTiO$_{3}$ and BaTiO$_{3}$ films using the taken from \cite{PERTSEV_1998}
strain and temperature dependencies of coefficients of functional (\ref{Fun1}%
) and examining separately the cases of transitions through the $ac$ and $r$
phases.

In the case of PbTiO$_{3}$ all these regimes are clearly visible and are
located inside the region of thermodynamically stable $c$-phase as shown in
Fig.~\ref{DIAG}~(a). To study the transient polarization dynamics we select
the representative points for each transition region (points A, B and C in
Fig.~\ref{DIAG}~(a)) and numerically solve the Landau-Khalatnikov kinetic
equations.%
\begin{equation}
L_{i}\frac{\mathrm{d}P_{i}}{\mathrm{d}t}=-\frac{\delta \tilde{G}}{\delta
P_{i}},
\end{equation}%
for each polarization component $P_{i}=P_{i}(t)$. Here $L_{i}$ are the
corresponding damping coefficients. The results are presented in Fig. \ref%
{COU} in form of experimentally measurable longitudinal and transversal
polarization currents $I_{l}=\frac{\mathrm{d}P_{3}}{\mathrm{d}t}$ and $I_{t}=%
\frac{\mathrm{d}P_{1}}{\mathrm{d}t}$.

Point A is selected for the $l$-switching region at $u_{m}=-0.0025$ and $%
T=500^{\circ }\mathrm{C}$. As it follows from Fig.~\ref{COU}~(a) the
polarization current has only the longitudinal component that is
characteristic for the longitudinal switching regime. Point B corresponds to
the $lt$-switching region and is taken at $u_{m}=-0.0025$, $T=250^{\circ }%
\mathrm{C}$. As shown in Fig.~\ref{COU}~(b) both components of polarization
current are observed but the transversal one is excited after the
longitudinal one and vanishes earlier than the longitudinal one. Point C is
taken in the \textit{\ }$t$-switching region at $u_{m}=-0.0025$ and $%
T=0^{\circ }\mathrm{C}$. As shown in Fig.~\ref{COU}~(c) the longitudinal and
transversal polarization currents are excited simultaneously.

In the case of BaTiO$_{3}$ [Fig.~\ref{DIAG}~(b)] one can observe only the $l$
and $t$-switching regimes. The $lt$-switching regime is difficult to detect
because of the very close location of the lines L$_{0}$ and L$_{1}$.

An important issue for the $t$-type switching is the dynamical
behavior of polarization just upon reaching the transversal
instability field. Under certain conditions the
intermediately-stable $ac$- or $r$- phase can be induced just above
$E^{(t)}$. Then, the continuous (as function of the field) turnover
of polarization through this phase will precede the abrupt
rotational drop-down. To distinguish between the shown in Fig.\thinspace \ref%
{DES} initially-continous ($t_{c}$) and straight-abrupt ($t_{a}$)
transversal switching process we study the global stability of functional (%
\ref{Fun1}) with respect to small deviations $\Delta P_{1}$, $\Delta P_{3}$
about the equilibrium point $P_{1}^{(t)}=0$ and $P_{3}^{(t)}=P_{3}\left(
E^{(t)}\right) $ exactly at $E=E^{(t)}$. This is a peculiar problem since at
$E=E^{(t)}$ the coefficient $H_{11}$ before $\left( \Delta P_{1}\right) ^{2}$
is equal to zero and the higher-order terms should be taken into account.
Following the catastrophe theory we keep only the most relevant terms and
present the expansion of (\ref{Fun1}) as:
\begin{equation}
\tilde{G}\approx \frac{1}{2}\rho \left( \Delta P_{3}\right) ^{2}+\mu \left(
\Delta P_{3}\right) \left( \Delta P_{1}\right) ^{2}+\frac{1}{2}\lambda
\left( \Delta P_{1}\right) ^{4},  \label{GG}
\end{equation}%
where $\rho =H_{33}$ (see (\ref{I2})), $\mu =\frac{1}{2}\frac{\partial ^{3}%
\tilde{G}}{\partial P_{3}\partial P_{1}^{2}}%
=b_{13}P_{3}^{(t)}+2b_{133}P_{3}^{(t)3}$ and $\lambda =\frac{1}{12}\frac{%
\partial ^{4}\tilde{G}}{\partial P_{1}^{4}}=\frac{1}{2}b_{33}$.
Transformation $x=\left( \Delta P_{1}\right) ^{2}$ and $z=\left( \Delta
P_{3}\right) $ maps the problem onto the study of quadratic functional $%
\frac{1}{2}\lambda x^{2}+\mu xy+\frac{1}{2}\rho y^{2}$. The last one
is globally unstable at $\lambda \rho >\mu ^{2}$ that provides the
straight-abrupt switching at $E\gtrsim E^{(t)}$. At $\lambda \rho
<\mu ^{2}$ this functional is locally stable and at the initial
stage of  reversal process the $P_{1}$ component develops
continuously as a function of the field. Using the given above
definition of $\lambda $, $\rho $, $\mu $ and excluding
$P_{3}^{(t)}$ according Eq. (\ref{Pc6}) we, after some algebra,
present the line L$_{2}$, separating these two regimes on the
$u_{m}$-$T$
phase diagram by equation:%
\begin{equation}
\text{L}_{2}:\quad PR=Q^{2}.  \label{L2}
\end{equation}%
with
\begin{eqnarray}
P &=&b_{3}b_{33}b_{13}-3b_{1}b_{33}^{2}+2b_{1}b_{13}^{2}-8b_{1}^{2}b_{133},
\label{PQR} \\
Q &=&5b_{1}b_{33}b_{333}-b_{3}b_{33}b_{133},  \notag \\
R
&=&3b_{33}^{2}b_{133}-2b_{13}^{2}b_{133}+8b_{1}b_{133}^{2}-5b_{13}b_{33}b_{333}.
\notag
\end{eqnarray}%
The $t_{c}$ and $t_{a}$ switching regions being located below and above this
line correspondingly.

Thoughtful analysis of equation (\ref{L2}) for BaTiO$_{3}$ case shows that
the line L$_{2}$ is located very close to the line L$_{1}$ (see Fig.~\ref%
{DIAG}~b) which means that "$t_{c}$-switching" always occurs
through the intermediate field-induced $ac$-phase. \ In contrast, the line L$%
_{2}$ does not exist in $c$-phase region of the phase diagram of PbTiO$_{3}$
[Fig.~\ref{DIAG}~(a)]  which implies the "$t_{a}$-switching"
through the intermediate $r$-phase takes place.

In this letter we have demonstrated the existence of different polarization
reversal regimes in strained pseudo-cubic ferroelectric PbTiO$_{3}$ and BaTiO%
$_{3}$ films. Depending on the temperature and on the misfit strain
one can distinguish the polarization reversal governed by the
longitudinal, transversal or mixed longitudinal-transversal
switching regimes. All three mechanisms can be observed in
PbTiO$_{3}$ compounds. In BaTiO$_{3}$ compounds only the
longitudinal  and transversal  mechanisms can be detected. The later
occurs through the intermediate $ac$-phase with initial continuous
turnover of the polarization vector as function of the field. The
dynamic appearance of the transversal polarization during transition
can be observed by the time-resolved piezo-force microscopy or by
the in-field Raman spectroscopy sensitive to the polarization vector
variation. Note, however that situation can be even more complex if
the 180$^{o}$ ferroelectric domains exist in the initial c-phase
or/and 90$^{o}$ transversal ferroelastic domains emerge during the
switching process. The study of such scenarios can be done on the
basis of presented above calculations.

This work was supported by FP7 ITN-NOTEDEV and IRSES-SIMTECH MC mobility
programs.


\newpage

\end{document}